# Depth of Field of Finite Energy Airy Shaped Waves

**K. Z. Nóbrega[1*], C. A. Dartora[2] and M. Zamboni-Rached[3]**

[1] *Federal Institute of Maranhão /Department of Electro-Electronics, São Luis-Ma, Brazil,*

[2] *Federal University of Parana/ Department of Electrical Engineering, Curitiba-Pr, Brazil,*

[3] *State University of Campinas/ School of Electrical and Computer Engineering, Campinas-Sp, Brazil,*

ABSTRACT – In the present paper the analytical aspects related to the maximum invariance depth of propagation finite energy Airy-type beams are investigated, considering not only the usual exponentially dumped Airy beams but also the truncated ones. It is obtained an analytic expression which seems capable of finding accurately the maximum invariance distance of these waves. The results seem to be in line with those obtained from numerical simulations.

## 1. Introduction

The most studied nondiffracting beams are certainly the Bessel and Mathieu Beams, for such reason their properties are very well known from both theoretical and experimental viewpoints. Other nondiffracting solutions are known to exist but their particular features are not fully explored yet. For instance, the Airy beams, which have attracted much attention due to their ability to accelerate and bend during propagation [1-4].

* Corresponding author : Federal Institute of Maranhão, Av. Getúlio Vargas 4, São Luis-Ma/Brazil. Tel.: +55 98 32189081. E-mail address: bzuza@ifma.edu.br .

The Airy wave solutions were first predicted by Berry and Balazs [5] within context of quantum mechanical non-relativistic Schrödinger's Equation, which possesses strong analogies with the paraxial wave equation in optics. Exploiting this analogy, Siviloglou *et. al.* obtained the analytical solution of Airy beams [1] in optics in 2007, being the experimental production obtained in the same year [6].

Theoretical studies of propagation's properties of finite energy Airy beams have been usually performed using exponentially dumped Airy beam [3], numeric simulations of finite aperture generation [7], and recently a simple and very efficient method for describing Airy-type beams truncated by finite apertures was developed [8] . Self-healing properties of Airy beams were also investigated both theoretically and experimentally in severe environmental conditions [4], confirming their robustness.

However, it is unusual to comment axial invariance of Airy beams as it was done in previous articles for Bessel beams [9, 10], for example. That is an important propagation property of any feasible beam specially because it sets a physical limit on its initial characteristics. As a fact, much of analysis found in literature related to Airy beams concerns only on their bending and accelerating characteristics. For those reasons, the maximum invariance distance of finite energy Airy beams will be investigated here, considering not only the usual exponentially dumped Airy beams but also truncated ones.

## 2. The Ideal Airy Beam

For the sake of completeness, a review of main theoretical aspects of Airy beams is given. As starting point, let us consider the paraxial wave equation, which is very often used in optics. In free space it is written as:

$$i\frac{\partial \psi}{\partial z} = -\frac{1}{2k}\nabla_{\perp}^{2}\psi ,\qquad(1)$$

where $\nabla_\perp^2$ is the transverse laplacian operator. In (1+1)D problems the field, $\psi$, depends only on the transverse coordinate $x$ and it propagates along $z$-axis, i.e., it is independent of coordinate $y$, making possible to write the transverse laplacian as $\nabla_\perp^2 = \partial^2/\partial x^2$. In this way, the formal solution of the paraxial wave Eq. (1) can be written, in integral form, as follows:

$$\psi(x,z) = \frac{1}{2\pi}\int_{-\infty}^{+\infty}\tilde{\psi}(k_x)\exp\left(i\frac{k_x^2}{2k}z + ik_x x\right)dk_x \qquad (2.1)$$

$$\tilde{\psi}(k_x) = \int_{-\infty}^{+\infty}\psi(x,0)\exp(-ik_x x)dx \qquad (2.2)$$

where $\psi(x,0)$ is the field distribution at $z = 0$ and $\tilde{\psi}(k_x)$ corresponds to its Fourier Transform. Them, if we follow the convention adopted in [1] and define two dimensionless variables $s = x/x_0$ and $\xi = z/(kx_0^2)$, it is possible to express the exponentially dumped Airy beam at $z = 0$ in the following way:

$$\psi(s,0) = Ai(s)\exp(as), \qquad (3)$$

where $a$ is the dumping parameter and $x_0$ is the spatial spot size. By itself, the Airy function $Ai(s)$ rapidly vanishes[2] for $s \to +\infty$, but for $s \to -\infty$ the decreasing is not fast enough to make it absolutely integrable. Thus, when an attenuating exponential factor $\exp(as)$ is introduced, it makes $\psi(s, 0) \to 0$ in the limit $s \to -\infty$ which assure that field will be absolutely integrable, i.e., it will have finite power flux. Therefore, if one takes Fourier's Transform of the previous equation using Eq. (2.2) in the normalized $k_x$-space, where $k_x' = k_x \cdot x_0$, one will easily obtain:

$$\tilde{\psi}(k_x') = \exp\left[\frac{i}{3}\left(k_x' + ia\right)^3\right]. \qquad (4)$$

---

[2] For $s \gg 0$, one has $Ai(s) \approx \dfrac{1}{2\sqrt{\pi}} \cdot s^{-1/4} \cdot \exp\left(-\dfrac{2}{3}s^{3/2}\right)$.

After a straightforward calculation, substituting Eq. (4) into Eq. (2.1), one finds:

$$\psi(s,\xi) = Ai\left[s - (\xi/2)^2 + ia\xi\right]\exp\left[\begin{array}{l}a(s - \xi^2/2) - i\xi^3/12 + \\ + i\xi/2(a^2 + s)\end{array}\right]. \quad (5)$$

The ideal Airy beam corresponds to the limit $a \to 0$ in Eq. (5), yielding

$$\psi(s,\xi) = Ai\left[s - (\xi/2)^2\right]\exp\left[i(\xi^3/12 + s\xi/2)\right]. \quad (6)$$

Looking at Eq. (6), corresponding to the ideal Airy Beam, one can clearly see its characteristics free of diffraction, infinity energy and bending along propagation. Related to the bending, notice that peak of intensity (maximum value of $|\psi(s,\xi)|^2$) follows the already known parabolic trajectory described by the expression below:

$$s = \frac{\xi^2}{4} + s_m , \quad (7)$$

where $s_m \cong -1.019$, which is the largest real root of $d/ds \cdot Ai(s) = 0$. Using the same numerical values of [1], it means, $\lambda$=0.5 µm and $x_0$=100 µm, it was found $\Delta x \approx 890$ µm (the beam's lateral deviation from the $z$ axis). Figure 1 illustrates the normalized field's intensity of an ideal Airy Beam where one could clearly see all those previous remarks mentioned about these waves.

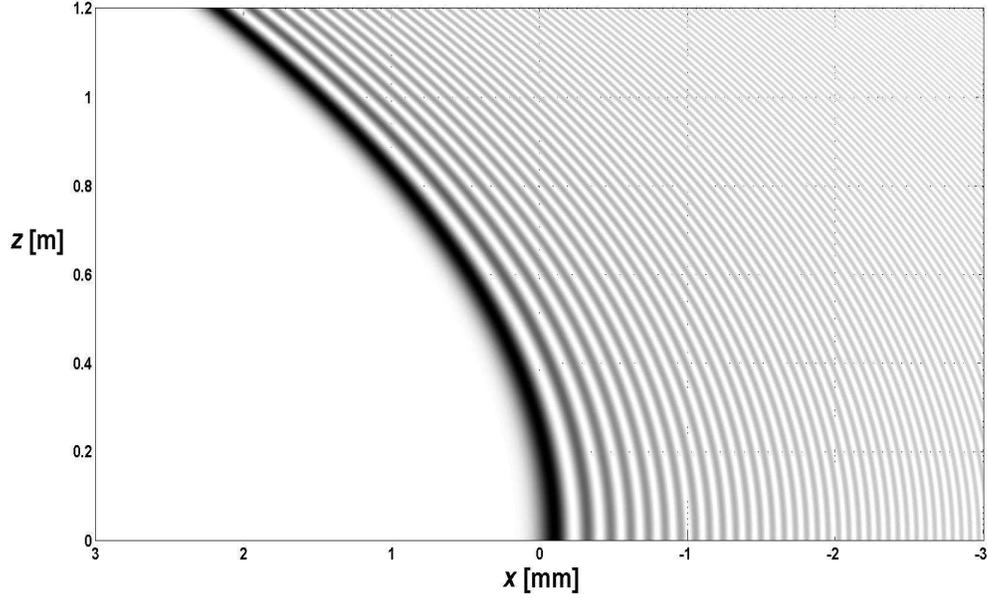

Figure 1. Propagation of an ideal Airy Beam considering $\lambda=0.5$ µm, $x_0=100$ µm.

## 3. Maximum invariance depth, $Z_{max}$

First of all, let us consider a finite energy Airy Beam of the form:

$$\psi(s,0) = Ai(s)\exp(as)u(s-s_\omega) , \qquad (8)$$

where $u$ (…) is the Heaviside Function and $s_\omega < 0$. An ideal Airy Beam purely truncated corresponds to the limit $a \to 0$ of the above expression, while a purely exponentially dumped Airy beam is obtained making $s_\omega \to -\infty$. For the case of purely truncated Airy beams ($a = 0$), the field $\psi(s, 0)$ abruptly goes to zero at $s = s_\omega$, contrasting to the exponentially dumped Airy beam described in the previous section, which smoothly goes to zero as $s \to -\infty$. Indeed, the truncated Airy beam can be thought as an ideal Airy beam partially blocked by an opaque wall which exists only for $s < s_\omega$.

As known from previous works related to Bessel Beams, when truncated by a finite aperture they propagate through a homogeneous medium keeping their characteristics of invariance only until a certain distance, $Z_{max} = R/\text{tg}(\theta)$ (where $R$ is the aperture radius and $\theta$ the cone angle of the Bessel beam). This distance characterizes the ability of these pseudo-nondiffracting beams to resist to diffractive effects only for a certain range which is, in general, beyond the usual diffraction length, $L_D=k.x_0^2$.

To obtain this depth of field, $Z_{max}$, in the case of finite energy Airy beams, we can use the properties of Fourier Transform, in particular the uncertain principle. The well-known uncertainty relation $\Delta\alpha.\Delta\beta \geq \frac{1}{2}$ is automatically implied by relations of Fourier Transform where $\Delta\alpha = \sqrt{\langle\alpha^2\rangle - \langle\alpha\rangle^2}$ and $\Delta\beta = \sqrt{\langle\beta^2\rangle - \langle\beta\rangle^2}$ are uncertainties in the spectral bandwidth of the function in $\alpha$ and $\beta$ spaces, respectively, and the average is defined as

$$\langle\alpha^n\rangle = \frac{\int_{-\infty}^{+\infty} \alpha^n |f(\alpha)|^2 \, d\alpha}{\int_{-\infty}^{+\infty} |f(\alpha)|^2 \, d\alpha} \tag{9}$$

Here, the field $\psi(x, z)$ is a function of the real space coordinates $(x, z)$ and can be transformed to the spatial frequency dual space, $(k_x, k_z)$, resulting in $\tilde{\psi}(k_x, k_z)$. Applying the uncertainty relation in the space $(z, k_z)$, it leads to $Z_{max} \approx 1/(2\Delta k_z)$, provided that one realizes $\Delta z = Z_{max}$. However, once that $\Delta k_z$ is related to the $(x, k_x)$ space through the paraxial equation so that one has $k_z = k_x^2/(2k)$, it yields to $\Delta k_z \approx k_x.\Delta k_x/k$ (of course this relation is valid for moderate values of $\Delta k_x$). Since $\Delta k_x$ is related to the aperture width $\Delta x$ by $\Delta x \approx 1/\Delta k_x$ one obtains

$$Z_{Max} \approx k \cdot \Delta x / \langle k_x \rangle_{rms}. \tag{10}$$

The above equation provides a very simple way of evaluating the depth of field of Airy-type beams with finite power flux. In the following, we are going to applied it to two interesting cases: the ideal and the exponentially dumped Airy beams truncated by a finite aperture.

## 3.1 The Ideal Airy truncated by a finite aperture

As previously mentioned, for a purely truncated Airy beams $a = 0$, $\Delta x = |s_\omega| x_0$ and the *rms*-value of $k_x$ is a nontrivial function of $|s_\omega|$, but in the range $|s_\omega| > 10$ it can be shown that $\langle k_x \rangle_{rms} = \gamma |s_\omega|^{1/2} / x_0$ where $\gamma = 0.57721$ is the Euler-Mascheroni constant. Therefore, after all those substitutions one can easily obtain the following maximum invariance distance:

$$Z_{max} \cong \gamma^{-1} |s_\omega|^{1/2} k x_0^2. \tag{11}$$

To the best of our knowledge, the above expression for purely truncated Airy beams was analyzed by our research group only. It allows us to estimate with good agreement the maximum distance where truncated Airy beams can propagate without suffer significant changes of their transverse intensity profile. Since that $k x_0^2$ is the usual diffraction length, the dimensionless $\xi_{max} = \gamma^{-1} |s_\omega|^{1/2}$ corresponds to the amplification of the maximum invariance distance with respect to usual diffraction limit ($L_D$), being a function of the normalized aperture $s_\omega$ only.

## 3.2 The exponentially dumped Airy beams truncated by a finite aperture

For the sake of comparison, for exponentially dumped Airy beams with $s_\omega \to -\infty$ the parameter limiting the maximum invariance distance is the attenuation constant $a$. Following the theoretical arguments presented above it can be shown that the invariance length of the exponentially dumped Airy beam is given by

$$Z_{max} \cong \frac{1.27}{\sqrt{a}} k x_0^2. \tag{12}$$

As can be seen below, in contrast to the case of truncated Airy beams, for which the maximum field intensity oscillates around the initial value until it abruptly decays to negligible values, the maximum field intensity of the exponentially dumped case, given by Eq. (12), smoothly decays as $\exp(-a\xi^2/2)$. Between these two extremes, it means, a truncated Airy beam and one exponentially dumped, there is an intermediary case corresponding to $-\infty < s_\omega < 0$ and $a > 0$, for which a conservative estimation of $Z_{max}$ can be given by the minimum between the values obtained from Eqs. (11) and (12).

Following, in Figure 2 it is shown the evolution of the maximum value of field's intensity as a function of the propagated distance $z$ for the purely Airy Beam truncated, exponentially truncated and exponentially dumped using some representative values of parameters $a$ and $s_\omega$. The values $\lambda = 500$ nm and $x_0 = 100\mu$m were kept fixed. Figure 2 shows evolution of peak intensity for the three different situations explained before. As expected, one can notice that maximum intensity belongs to the truncated Airy beam with fast oscillations as the propagation distance increases (consequence of the sharp edge diffraction occurring at $x = s_\omega . x_0$), until it abruptly decays to negligible values.

It happens because that field keeps the same airy properties except for truncation, and its oscillations are similar to those seen on truncated Bessel and Mathieu beams. On the contrary, the exponentially dumped Airy beam does not present oscillations, reflecting the soft decaying of the field intensity at $z = 0$ resembling the behavior of other softly attenuated finite energy beams such as Bessel-Gauss beams. For the truncated exponentially dumped Airy beam the fast oscillations are mounted over the envelope in the form $\exp(-a\xi^2/2)$.

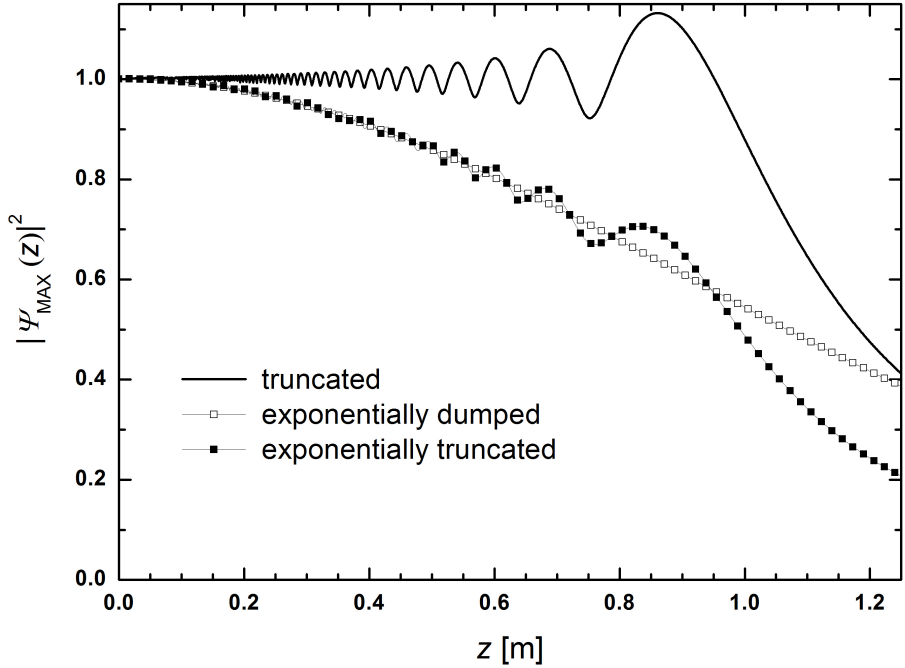

Figure 2. Evolution of the maximum field intensity, $|\Psi_{max}(z)|^2$, as a function of propagated distance $z$ for truncated ($s_\omega = -30$), exponentially dumped ($a=0.02$) and exponentially truncated ($a=0.03$ and $s_\omega = -30$) Airy beams considering $\lambda=0.5$ µm, $x_0=100$ µm.

In Figure 3 it is described the transverse shift of the maximum field intensity from the $z$ axis along the propagation of an ideal Airy beam, given by $\Delta x = z^2/(4x_0^3 k^2)$, and a comparison of it to the shifting of those three different Airy-type beams previously discussed. Looking at the figure, one clearly sees that along the propagated distance, the shifting of the truncated wave is almost similar to the ideal case, which shows how strong the Airy characteristics are still present in that wave. However, this is not evident for the other two waves because one could realize that for a certain distance, $z$, there is an abrupt changing of behavior which means on that point the wave is far away from "Airy" behavior. Indeed, it looks like a kurtic distribution. In general and under same circumstances, that figure also turns clear how an exponentially dumped keeps for a longer range its characteristics of Airy function in comparison to exponentially truncated once that $\Delta x$ abruptly changes for a higher distance.

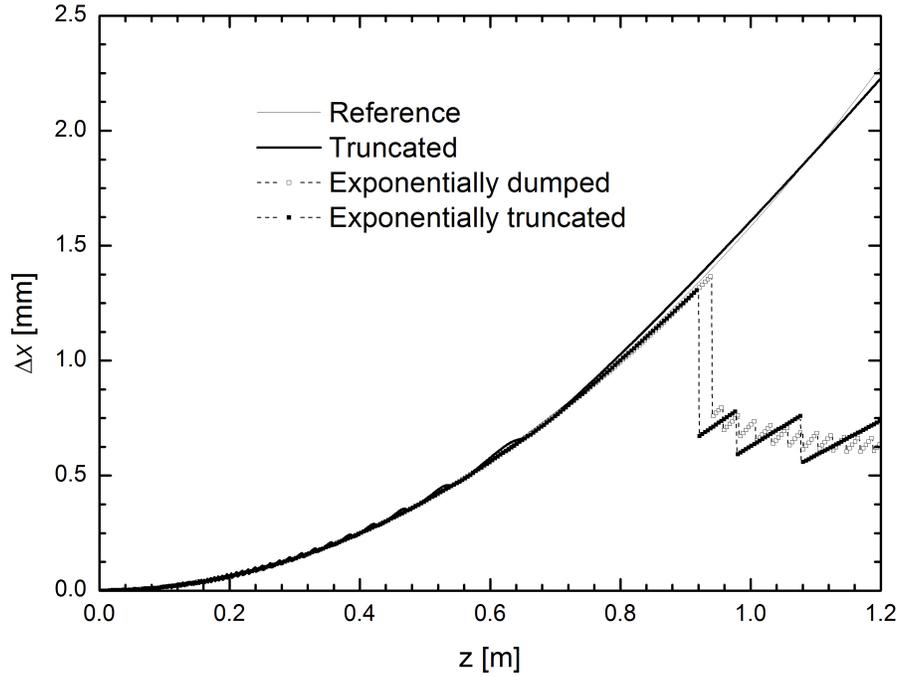

Figure 3. Beam shift $\Delta x = x - s_m x_0$ as a function of propagated distance for ideal Airy Beam (reference) comparing to truncated ($s_\omega = -5$), exponentially dumped ($a=0.1$) and exponentially truncated ($a=0.1$ and $s_\omega = -5$). Wavelength and spot size are kept the same.

Figures 4 and 5 compare numerical simulations and analytical results of $Z_{max}$ obtained through Eq. (11) for truncated Airy as function of $s_\omega$, respectively. As a fact, in Figure 4, two different spot sizes have been considered and one can notice the same behavior independent of $x_0$ chosen. Actually, that result is totally understood in Figure 5, where the error between analytical formulae and numerical results are compared, showing that there is a true association between numerical results (here taken as reference) and formulae proposed. It can be understood if one takes account only one single value of $s_\omega$ and to observe that no difference appears for different spot sizes. Considering how the error has been calculated, $\left(\dfrac{Z_{Max}^{Ana} - Z_{Max}^{Num}}{Z_{Max}^{Num}}\right).100$, it could happen only if a complete match between both $Z_{Max}$ concerning variation with $x_0$ parameter.

Still related to Figure 5, one can notice a high error for $s_\omega$ values smaller than ten. That is understandable because on that region the Airy beam would be almost completely cut off from its "airy" characteristics (and, furthermore, the relation $\Delta k_z \approx k_x \cdot \Delta k_x / k$ loses its validity). In other words, in that situation one is claiming Airy beam propagation but beams shape does not have such correspondence. In general, there is a good agreement between theoretical predictions and the actual values found, showing that the obtained expression can give valuable insights into the propagation characteristics of finite energy Airy beams.

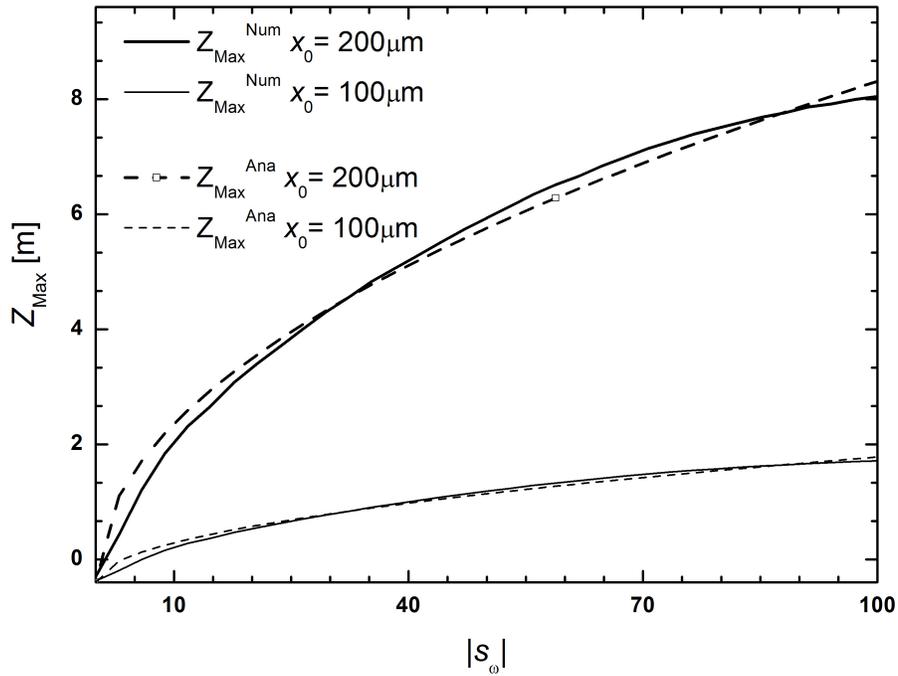

Figure 4. Comparison between numerical simulations and the analytical formulae (11) to find $Z_{max}$ for a truncated Airy Beam as function of $|s_\omega|$ with two different spot sizes.

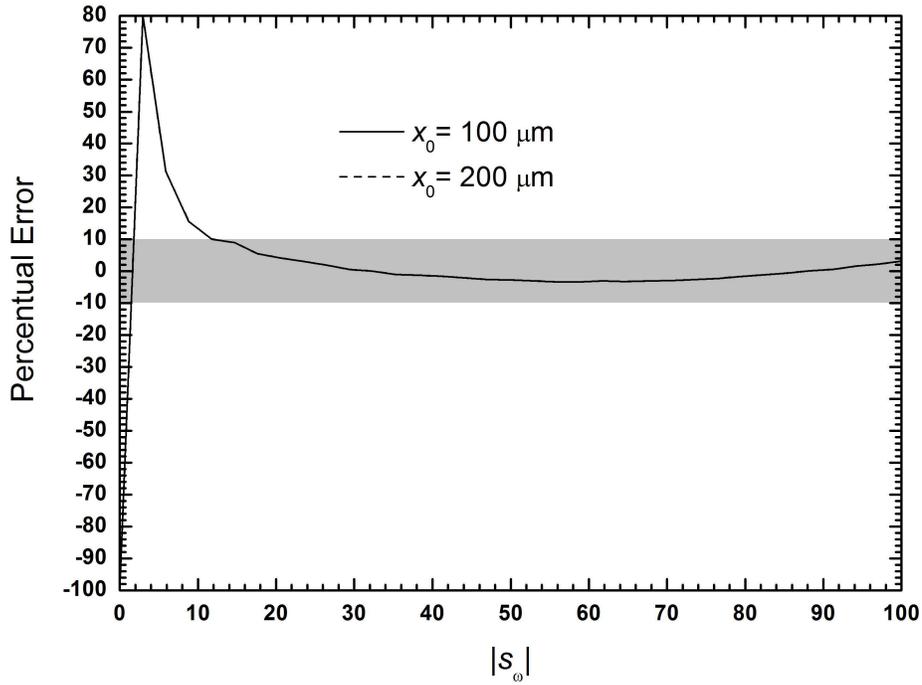

Figure 5. Percentual error for situation analyzed in Figure 4 considering both spot sizes.

Figures 6 and 7 show the same previous analysis but considering Eq. (12), i.e., comparison between numerical and analytical results of $Z_{Max}$ but for exponentially dumped Airy beam. Looking at Eq. (12) the difference is the parameter involved, it means, instead of $Z_{Max}$ to be defined as a function of $s_\omega$ it will be written using $a$. Different values of $a$ have been considered for two different spot sizes. As seen on Figure 4, Figure 6 shows similar results related to general behavior of exponentially dumped Airy beam considering two different spot sizes. Additionally, in Figure 7 one can identify only one error curve predicting the same conclusions related to Figure 5, claiming that a correct match between analytical formulae proposed and numerical results is reasonable.

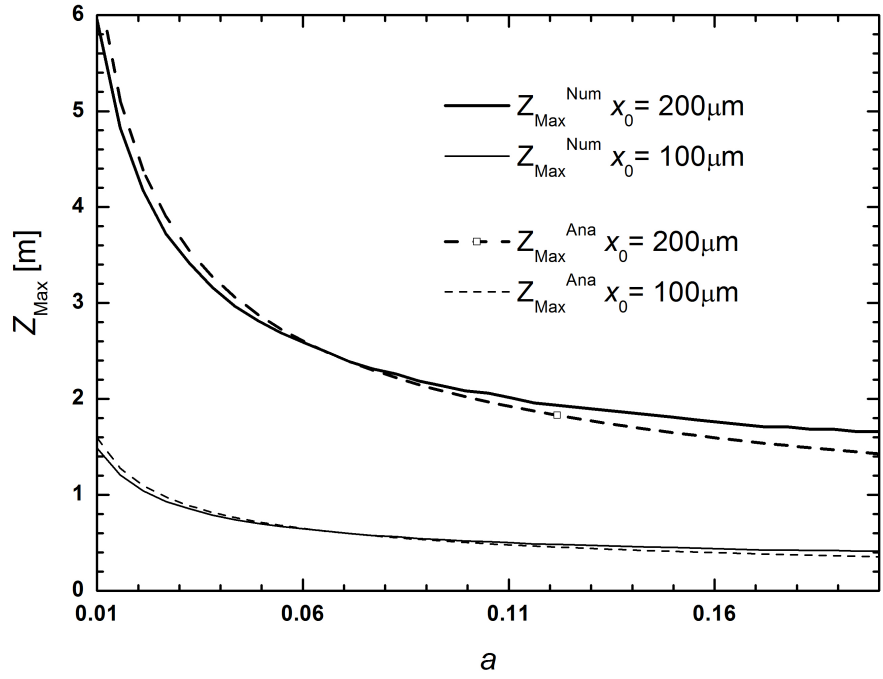

Figure 6. Comparison between numerical simulations and the analytical formulae, Eq. (12), to find $Z_{max}$ for exponentially dumped Airy Beam as function of $a$. Two spot sizes were considered.

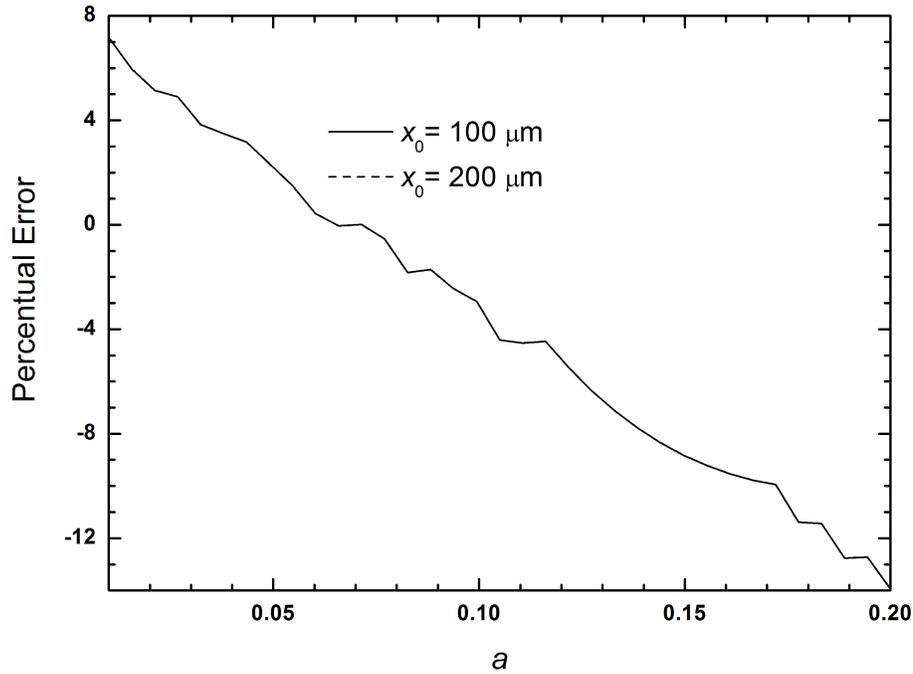

Figure 7. Percentual error for situation analyzed in Figure 6 considering both spot sizes.

# 4. Conclusions

In this work it was discussed features of ideal Airy beams and their ability to accelerate and to bend along propagation. Besides that, truncated Airy Beams under paraxial regime were also discussed showing these waves preserve resemblances with other truncated nondiffracting beams, such as Bessel and Mathieu beams. In fact, a deeper analysis was focused on finding $Z_{max}$ as a simple formula able to describe it for both truncated and exponentially dumped Airy beams, which are the most discussed Airy type waves in literature. Finally, analytic expressions for the maximum invariance distance were compared with numerical values found through simulation of paraxial wave equation, showing good agreement as predicted.

# Acknowledgements

This research has been supported with funding from agencies FAPESP (grant 11/51200-4), CNPq (grant 307962/2010-5), FAPEMA (Process number 002244/2012) as well as IFMA (PRPGI/23249.025881.2012/14). Authors, are also in debt with DEE/UFPr and DMO/UNICAMP.

**Figure Captions**

Figure 1. Propagation of an ideal Airy Beam considering $\lambda$=0.5 µm, $x_0$=100 µm.

Figure 2. Evolution of the maximum field intensity, $|\Psi_{max}(z)|^2$, as a function of propagated distance $z$ for truncated ($s_\omega$= -30), exponentially dumped ($a$=0.02) and exponentially truncated ($a$=0.03 and $s_\omega$= -30) Airy beams considering $\lambda$=0.5 µm, $x_0$=100 µm.

Figure 3. Beam shift $\Delta x = x - s_m x_0$ as a function of propagated distance for ideal Airy Beam (reference) comparing to truncated ($s_\omega$= -5), exponentially dumped ($a$=0.1) and exponentially truncated ($a$=0.1 and $s_\omega$= -5 ). Wavelength and spot size are kept the same.

Figure 4. Comparison between numerical simulations and the analytical formulae (11) to find $Z_{max}$ for a truncated Airy Beam as function of $|s_\omega|$ with two different spot sizes.

Figure 5. Percentual error for situation analyzed in Figure 4 considering both spot sizes.

Figure 6. Comparison between numerical simulations and the analytical formulae, Eq. (12), to find $Z_{max}$ for exponentially dumped Airy Beam as function of $a$. Two spot sizes were considered.

Figure 7. Percentual error for situation analyzed in Figure 6 considering both spot sizes.